\documentclass[twocolumn]{openjournal}

\usepackage{xcolor}
\usepackage{textgreek}
\usepackage[utf8]{inputenc}
\usepackage[english]{babel}
\usepackage{subfig}
\usepackage{natbib}
\usepackage{hyperref}

\hypersetup{
    unicode, 
    colorlinks=true,
    linkcolor=linkcolor,
    citecolor=linkcolor,
    filecolor=linkcolor,
    urlcolor=linkcolor,
}

\usepackage{color,colortbl}
\definecolor{linkcolor}{rgb}{0.0,0.3,0.5}
\usepackage{tensind}
\tensordelimiter{?}
\DeclareGraphicsExtensions{.bmp,.png,.jpg,.pdf}
\usepackage{verbatim}
\usepackage[normalem]{ulem}
\usepackage{orcidlink}
\usepackage{soul}

\newcommand{\umaiii}{UMa III/U1}
\newcommand{\kpc}{\mathrm{kpc}}

\newcommand{\msun}{\mathrm{M_\odot}}
\newcommand{\lsun}{\mathrm{L_\odot}}

\graphicspath{ {./figs/} }

\begin{document}
\title{A Bayesian Exploration of the Mass of Ursa Major III: \\ 
Kinematics, Rotation and their influence on the Mass to Light Ratio}

\author{{Tim R. Adams$^{1}$\orcidlink{0009-0006-2556-9983}}, 
{Brendon J. Brewer$^{2}$\orcidlink{0000-0001-9902-7112} \& Geraint F. Lewis$^{1}$\orcidlink{0000-0003-3081-9319}}}

\affiliation{$^{1}$Sydney Institute for Astronomy, School of Physics, A28, The University of Sydney, NSW 2006, Australia \\ $^{2}$Department of Statistics, The University of Auckland, Private Bag 92019, Auckland 1142, New Zealand}

\email{tada6448@uni.sydney.edu.au}
\email{bj.brewer@auckland.ac.nz}
\email{geraint.lewis@sydney.edu.au}

\begin{abstract}

We investigate the kinematics of the potential ultra-faint dwarf galaxy (UFD) \umaiii\ using Bayesian inference to search for the signal of any potential intrinsic rotation. The magnitude of rotation is relevant to estimating the total mass of \umaiii\, which is critical in determining whether or not \umaiii\ is in fact a UFD, or possibly a star cluster home to a significant binary fraction. A non-rotating model and a rotational model are fitted for the current total population of member stars of \umaiii\, finding that a non-rotating model was preferred by a factor of $\sim5-12 \times$. This was repeated on a reduced population of \umaiii\, where potential contaminant stars were removed. A similar preference for non-rotation was found for these reduced populations. We calculate a lower-bound rotational mass estimate for \umaiii\ and a corresponding lower bound mass-to-light ratio of $734.4^{+339.0}_{-176.2} \, \msun/\lsun$ for the total population. We conclude that \umaiii\ still remains an ambiguous object with viable arguments for both the UFD and self-gravitating star cluster scenarios, however under both, \umaiii\ is unlikely to be supported by rotational pressure.

\end{abstract}


\maketitle

\section{Introduction}
\label{sec:intro}

Since the discovery of the Sculptor galaxy --- the first faint dwarf galaxy of the Milky Way --- the search for similar systems has widened, and as deeper field surveys became available, fainter systems have been observed \citep{1938BHarO.908....1S}. However, it would take until 2005 for the first ultra-faint dwarf galaxies (UFDs) to be identified by the Sloan Digital Sky Survey I (SDSS;  \citet{Willman_2005a, Willman_2005b}). UFDs are defined to have absolute V-band magnitudes of $ M_V \geq -7.7$ (which corresponds to $L \leq 10^5 \, \lsun$) and are among the oldest and most metal-poor stellar systems known \citep{2019ARA&A..57..375S}. The first spectroscopic studies on UFDs were done by \citet{Kleyna_2005} and \citet{Mu_oz_2006} on UMa 1 and Bo\"{o}tes 1 respectively, and concluded that UFDs could not be purely baryonic systems. Follow up work by \citet{Martin_2007} and \citet{Simon_2007} using the Keck II/DEIMOS instrument on remaining UFDs at the time further strengthened this conclusion. Kinematic analyses of known UFDs have revealed a discrepancy between the dynamical and stellar mass of these systems, implying the existence of significant amounts of dark matter in each UFD \citep{2019ARA&A..57..375S}. 

These characteristics have led UFDs to become promising locations to probe the nature of dark matter on small scales as well as understand the processes that govern how the first galaxies form \citep{Bovill_2009,Bovill_2011,Calabrese_2016,Errani_2018}. The total number of UFDs in the Milky Way also places an upper bound on the lowest mass a halo can have, which constrains the mass of the dark-matter particle \citep{10.1093/mnras/stx2330}.

Probing towards the faintest end of the galactic luminosity function does not come without its challenges. At a brightness of $M_V < -5$, a clear distinction can be made between dwarf galaxies and star clusters by looking at their corresponding half-light radius. At this brightness, objects with $r_h < 20 \, \rm pc$, can be identified as globular clusters whereas above $r_h  > 100 \, \rm pc$ the objects can be confidently identified as dwarf galaxies. However, at lower brightnesses it becomes more difficult to classify a system purely based on its photometric information. This is due to the similar size distributions that dwarf galaxies share with star clusters at this brightness, particularly globular clusters \citep{2019ARA&A..57..375S}. Although dwarf galaxies and star clusters share some similarities, they differ substantially in the amount of dark matter contained within their half-light radius. Star clusters are known to have negligible amounts of dark matter, which is typically quantified as a mass-to-light ratio between $1.4  < M/L_V \,  (\msun/\lsun) < 2.5 $ \citep{2020PASA...37...46B}. In contrast, dwarf galaxies, and particularly UFDs, have a significantly larger proportion of dark matter, leading to mass-to-light ratios often exceeding $M/L_V \sim 10^3 \, \msun/\lsun $ \citep{2019ARA&A..57..375S}. The recently discovered Ursa Major III, known as \umaiii\ (UMa3/U1), is a satellite of the Milky Way \citep{2024ApJ...961...92S}. Located at a heliocentric distance of $~10 \, \kpc$, this ambiguous object is notable for its extremely low luminosity ($M_{\rm V} = +2.2^{+0.4}_{-0.3}$), minimal stellar mass ($M_{\rm tot} =  16^{+6}_{-5} \, \msun$) and small size ($r_{\rm h} = 3 \pm 1 \, \rm pc$), containing only about 60 stars \citep{2024ApJ...961...92S,2024ApJ...965...20E}. It therefore presents a unique opportunity to study the faintest known satellite, and in the case that it proves to be a dwarf galaxy, \umaiii\ would offer insights into the formation and evolution of dwarf galaxies in the universe \citep[e.g.][]{2024ApJ...968...89E}. \umaiii\ may also have one of highest mass-to-light ratios for Milky Way dwarf galaxies, estimated to be $ M/L_V \sim 6500 \, \msun/\lsun$, substantially higher than that of typical galaxies and suggesting that \umaiii\ contains a substantial amount of dark matter \citep{2024ApJ...961...92S}. This mass-to-light estimation was done through a kinematic analysis of member stars of \umaiii\ \citep{2024ApJ...961...92S}. \citet{2024ApJ...961...92S} note, however, that these initial mass-to-light ratios should be cautiously interpreted as there are still unknowns in the data such as potential contaminant binary systems and tidal disruption effects. Nonetheless, the potential presence of such a high proportion of dark matter relative to its visible matter makes \umaiii\ an important object of study for understanding the nature and distribution of dark matter in the universe. It also raises questions about the processes that lead to the formation of such dark matter-dominated systems \citep{Bullock_2017, Battaglia2022, sales2022baryonicsolutionschallengescosmological}.

Previous estimates of the mass-to-light ratio of \umaiii\ used a non-rotating kinematic model to predict velocities of member stars \citep{2024ApJ...961...92S}. This in turn assumes that \umaiii\ is completely supported by dark matter pressure as opposed to rotation, which has a direct influence on the estimated velocity dispersion of the system, and in turn the mass-to-light ratio \citep{2010MNRAS.406.1220W}. Given how important the mass-to-light ratio estimate is to various other quantities of \umaiii, further kinematic exploration is needed. The primary objective of this study is to determine whether \umaiii\ exhibits any rotation and to what extent this affects the estimated mass-to-light ratio.

This paper is structured as follows: First, in section ~\ref{sec:background} we summarise the current literature on \umaiii. In Section~\ref{sec:approach} we outline our approach, detailing the data in Subsection~\ref{subsec:data}, our Bayesian approach in~\ref{subsec: bayesianinference} and our kinematic modelling in~\ref{subsec: kinematicanalysis}. We present our results in~\ref{sec:results} and our conclusions in~\ref{sec:conclusions}.

\section{Background/Literature Review}
\label{sec:background}

The potentially high mass-to-light ratio of \umaiii\ has broader implications for cosmology and galaxy formation theories. In the Lambda Cold Dark Matter (ΛCDM) paradigm, numerous low-mass dark matter halos are predicted to exist that may host faint galaxies like \umaiii\ \citep{Klypin_1999,Bovill_2009,Bullock_2017,2019ARA&A..57..375S}. Studying \umaiii\ may help refine these formation models and improve our understanding of the minimum halo mass required for galaxy formation. It also provides a valuable test case for alternative dark matter theories, which could offer new insights into the fundamental properties of dark matter itself \citep{2024PhRvD.110j3048Z,2024ApJ...968...89E}. Additionally, \umaiii\ may be a dwarf spheroidal galaxy (dSph), which makes it an attractive candidate for studying dark-matter annihilation. Although \umaiii\ is at the lower end of dSph masses, its close proximity and potentially high mass-to-light ratio lead to a J-factor of $\sim 10^{21}\ \rm GeV^2s^{-5}$, distinctly greater than the commonplace value of $\sim 10^{20}\ \rm GeV^2s^{-5}$ \citep{zhao2024}.

\umaiii\ was identified by the Ultraviolet Near Optical Northern Survey (UNIONS) in April of 2023 and first presented by \citet{2024ApJ...961...92S}. A population of 11 member stars were identified through a combination of photometry, Gaia astrometry, and spectroscopic follow ups with the Keck/DEIMOS instrument.

Through kinematic analysis, \citet{2024ApJ...961...92S} show that a velocity dispersion of $\sigma = 3.7 \, \rm km\,s^{-1}$ best fit the data of 11 radial velocity members. Exclusion of one of these stars, however, sees this value drop to $1.9 \, \rm km\,s^{-1}$, and becomes unresolved when another member is excluded. This sees the mass-to-light ratio drop from  $\sim 6500 \, \msun/\lsun$ to $\sim 1900 \, \msun/\lsun$, and then undetermined. This highlights the need for repeat velocity measurements of the 11 radial members and further kinematic analysis of the population.

\citet{2024ApJ...965...20E} highlights the sensitivity of the mass-to-light ratio to the chosen member stars and notes that even though the mass-to-light ratio may not be well resolved, a genuine dark-matter dominated UFD is still a possible description of \umaiii. They show this through analysis of simulations under two set ups; the first assuming that \umaiii\ is in-fact a self-gravitating star cluster with little to no dark matter, and then looking at a dark matter dominated model. They show that should \umaiii\ be a self-gravitating star cluster, it would have an average density similar to that of the average density of the Milky Way at the pericenter of its orbit. Therefore in this scenario it is extremely unlikely that \umaiii\ could resist tidal stripping from the Milky Way for any reasonable time given its short orbital period. It’s also important to note that in the self-gravitating set up, a total stellar mass of 16 $\msun$ would lead to a line-of-sight velocity dispersion of $\sigma_{\rm LOS} \sim 50 \, \rm m\,s^ {-1}$, significantly smaller than the $3.7 \, \rm km\,s^{-1}$ found in the previous kinematic analysis \citep{2010MNRAS.406.1220W}. \citet{2024ApJ...965...20E} then investigated a dark matter dominated model, showing that a much higher average density of \umaiii\ is achieved (roughly x$1000$ that of the average density of the Milky Way), protecting \umaiii\ from tidal stripping effects. Importantly, this estimated averaged density of \umaiii\ would make it the densest UFD ever detected, if it is in fact a UFD.

\citet{devlin2025reevaluatinguma3u1starcluster} revisited the dynamical cluster simulations done by \citet{2024ApJ...965...20E} using a collisional N-body software, NBody7. \citet{devlin2025reevaluatinguma3u1starcluster} show that, through the inclusion of primordial binaries in their simulations, a luminosity weighted velocity dispersion of $\sigma = 4.75 \, \rm km\,s^{-1}$ can be recovered. This value is consistent with the $\sigma = 3.7 \, \rm km\,s^{-1}$ found by \citet{2024ApJ...961...92S}, and in disagreement with that found by \citet{2024ApJ...965...20E}. \citet{devlin2025reevaluatinguma3u1starcluster} argue that the discrepancy in the estimated velocity dispersion is due to the omission of compact remnants in the simulations done by \citet{2024ApJ...965...20E}. The omission of these compact remnants (which make up roughly 50\% to 80\% of the cluster mass in the simulations done by \citet{devlin2025reevaluatinguma3u1starcluster}) leads to a significant underestimation in the total cluster mass. \citet{devlin2025reevaluatinguma3u1starcluster} argues that the collisionless numerical approach used by \citet{2024ApJ...965...20E} can't reproduce individual stellar interaction, and effects such as mass segregation are thus ignored. This leads to an underestimation in the cluster density, particularly in the core of the cluster. The inclusion of collisional interactions and primordial binaries in their cluster simulations leads to \citet{devlin2025reevaluatinguma3u1starcluster} concluding that \umaiii\ could in fact be a star cluster, with an average remaining lifetime between $1.9-2.7$ Gyr. This is significantly greater than the $0.4$ Gyr that \citet{2024ApJ...965...20E} had estimated, indicating that the observation of \umaiii\ as a star cluster would not have to be so ``fine-tuned'' as \citet{2024ApJ...965...20E} remarks.

Additional N-body simulations of \umaiii\ done by \citet{Rostami_Shirazi_2025} showed that a dark matter deficient progenitor most accurately reproduced the current day properties of \umaiii, further supporting the case that \umaiii\ is a self-gravitating star cluster. Follow up spectroscopic analysis by \citet{cerny2025} favoured small velocity and metallicity dispersions, as well as setting upper limits on both values. These values do not provide observational evidence for dark matter and instead favour a star-cluster population - while a UFD scenario remains viable pending high-resolution abundance patterns and even tighter multi-epoch kinematics.

\section{Approach}
\label{sec:approach}

\subsection{Data}
\label{subsec:data}

\begin{figure}[t!]
    \centering
    \includegraphics[width=8.5cm]{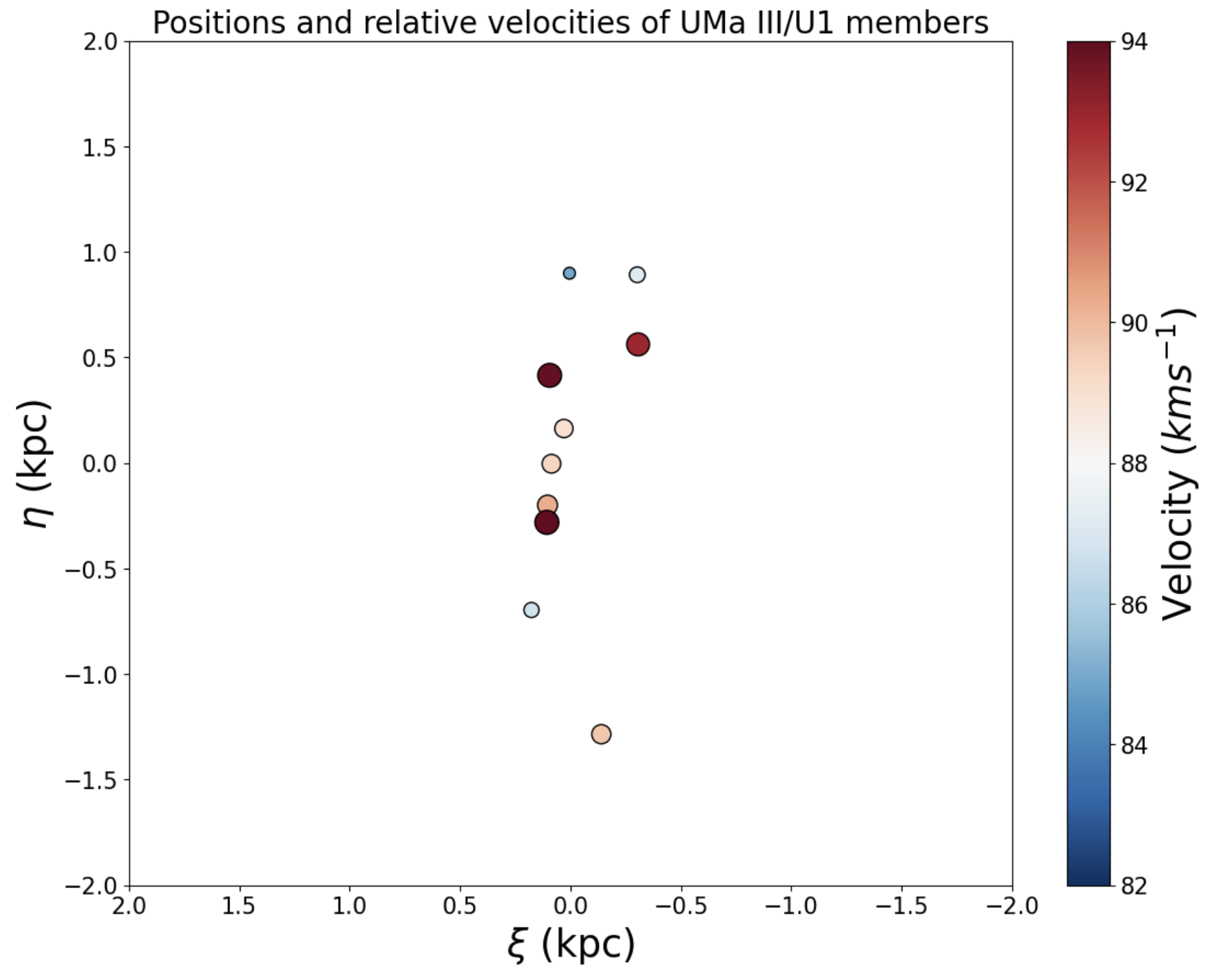}
    \caption{Positions and relative velocities of the $11$ radial velocity members of \umaiii. The size of each marker corresponds to the member velocity.}
    \label{fig:positions}
\end{figure}

All data used in this analysis was obtained from Table 3 in the original discovery paper by \citep{2024ApJ...961...92S}. The $11$ radial velocity members were identified as a population through the use of a matched filter algorithm. A detailed explanation of how membership likelihoods were assigned and the matched filter algorithm can be found in the original discovery paper by \citet{2024ApJ...961...92S}. All $11$ member stars are displayed in Figure \ref{fig:positions}.

\subsection{Bayesian Inference}
\label{subsec: bayesianinference}

To detect possible rotation in \umaiii, Bayesian inference will be used to determine the posterior distributions of parameters belonging to a rotational model (shown in Table \ref{table:params}). Studying these posterior distributions will allow us to quantify the probability that \umaiii\ is rotating, and consequently estimate the total mass of \umaiii. These posterior distributions are given by Bayes's theorem:

\begin{equation}
    P(\omega|D,I) = \frac{P(\omega|I)P(D|\omega,I)}{P(D|I)}
\end{equation}

\noindent Here $\omega$ is a vector of model parameters, $D$ is the given data (the velocity measurements for radial members) and $I$ is the prior information we 
assume, such as the choice of the exact form of the kinematic model.
We can think of $P(\omega|D,I)$ as a measure of the plausibility of 
different values of the unknown parameters $\omega$, given a set of data
$D$ and some prior information $I$ \citep{Gregory_2005}.
$P(\omega|I)$ is the prior distribution for the model parameters, describing the assumed uncertainty based on the prior information $I$ alone. Once we receive some new data $D$, the likelihood function $P(D|\omega,I)$ can then be calculated, which represents the probability of obtaining that data as a function
of the unknown parameters $\omega$. The final term $P(D|I)$ (called the
evidence or the marginal likelihood)
behaves as a normalisation factor. Bayes's theorem describes a type of learning process, by which the probability distribution of the model parameters can be updated from the prior to the posterior distribution
in order to take into account the data $D$.

This approach also naturally leads to a way of comparing two models $M_1$ and $M_2$,
with potentially different parameter spaces,
through the Bayes factor, $K$. The Bayes factor is defined as:

\begin{equation}
    K(M_1,M_2) = \frac{P(D|M_1,I)}{P(D|M_2,I)}
\end{equation}

\noindent and can be thought of as a likelihood ratio between $M_1$ and $M_2$, i.e, if $P(D|M_1,I)  = \, 2P(D|M_2,I)$, then the data is twice as likely under $M_1$ compared
to $M_2$. Posterior model probabilities can also be calculated. 

When we compute the respective Bayesian evidence, $P(D|M,I)$, and parameter posteriors for each model, the nested sampling implementation UltraNest was used \citep{2021JOSS....6.3001B}. UltraNest allows us to obtain the desired posterior model parameter distributions and evidence. The evidence integral for a model $M$ is given by:

\begin{equation}
    \mathcal{Z} = P(D|M) = \int P(D|\omega, M) P(\omega|M) \, d\omega
\end{equation}

\noindent i.e., the integral of the prior distribution times the likelihood function, over the entire parameter space of the model.

\subsection{Kinematic Analysis}
\label{subsec: kinematicanalysis}

\begingroup 
    \setlength{\tabcolsep}{9pt} 
    \renewcommand{\arraystretch}{1.5} 
    \setlength\extrarowheight{2pt}
    \begin{table}
        \centering
        \caption{\small Prior distributions for $V_1$ parameters.
            For the Student-$t$ distribution, the parameters are the location,
            scale, and shape respectively.}
        \begin{tabular}{ c c c }  
            Parameter & Description & Prior\\
            \hline     
            $ \log_{10}\,\sigma_{\rm v}$ & Velocity dispersion & $ \rm Student(1,0.5,4)$\\
            $A$ & Amplitude & $10^{\rm Student(1,0.5,2)}\sigma_{\rm v} $\\  
            $\phi$ & Angle of rotation &  $ \rm Uniform(0, 2 \pi )$\\  
            $v_{\rm sys}$ & Systemic velocity & $ \rm Student(88.6, 0.1\sigma_v \,, 1)$\\  
            \hline  
        \end{tabular}  
        \label{table:params}
        \vspace{13pt}
    \end{table}
\endgroup

To describe the velocity distribution of \umaiii\ member stars, the $V_1$ single component rotational model prescribed by \citet{Veljanoski_2014} is used. This model consists of a rotational component and a velocity dispersion. The full line-of-sight velocity model can be written as:

\begin{equation}\label{eq:vlos}
    v_{\rm LOS} (\theta) = v_{\rm rot}(\theta) + v_{\rm sys}
\end{equation}

\noindent With the rotational component as,

\begin{equation}\label{eq:vrot}
    v_{\rm rot} (\theta) = A\sin(\theta - \phi)
\end{equation}

\noindent Where $A$ is the rotational amplitude, $\theta$ is the on-sky angle of each star in plane-polar coordinates, $\phi$ is the angle of rotation and $v_{\rm sys}$ is a systemic velocity. If the measured velocities are assumed to be normally distributed with a mean value of $v_{\rm LOS}$ and dispersion $\sigma$ given by:

\begin{equation}\label{eq:disp}
    \sigma^2 = (\Delta v)^2 + \sigma^2_v
\end{equation}

\noindent Where $\Delta v$ is the error in the line-of-sight velocity, and $\sigma_v$ is the velocity dispersion. Under this assumption, the likelihood function (log-likelihood is used in practice) becomes:

\begin{equation}\label{eq:logL}
    \ln \mathcal{L}_M  =  -\frac{N}{2}\ln{2\pi\sigma^2} -\frac{1}{2}\sum_i \frac{(v_i - v_{\rm LOS})^2}{\sigma^2}
\end{equation}

\noindent Here $N$ is the number of members and $v_i$ is the line-of-sight velocity for the $i^{th}$ member.

\subsection{Priors}

For this analysis, the choice of prior distributions will be done in a similar manner to that by \citet{2024PASA...41...73L}, whereby a more complex prior distribution is constructed to restrict the rotational amplitude to being either negligible, moderate or large, with roughly equal probabilities. Student's t-distributions are chosen over Gaussian distributions here as their heavier tails allow us to account for any biases in our priors. The defined prior distributions are shown in Table \ref{table:params}. This is done by assuming that the rotational amplitude, $A$, and systemic velocity, $v_{ \rm sys}$, are functions of $\sigma_v$. The systemic velocity prior is centered around $v = 88.6 \, \rm km\,s^{-1}$, which is based on the results from \citet{2024ApJ...961...92S}.
The prior median for the velocity dispersion is 10 kms$^{-1}$, with about 88\% of the probability contained between 1 and 100 kms$^{-1}$.
The prior median for $A$ is set to be 10 times the velocity dispersion, with
88\% of the probability contained between 0.1 and 100 times the velocity dispersion.
Both of these priors have heavy tails to account for any biases.
A sample of $10^5$ chains is run for each of the models.

\section{Results}
\label{sec:results}


\begingroup 
    \setlength{\tabcolsep}{9pt} 
    \renewcommand{\arraystretch}{1.5} 
    \setlength\extrarowheight{2pt}
    \begin{table}
        \centering
        \caption{\small Summary of the model parameter statistics for the total population $V_1$ model. The values shown are the median values of the posterior distributions. The ranges displayed cover a $68\%$ credible interval}
        \begin{tabular}{ c c c }  
            Parameter & Estimate & Units\\
            \hline     
            $ \sigma_{ \rm v}$ & $3.28^{+1.18}_{-0.89}$ & $ \rm km\,s^{-1}$\\
            $A$ & $0.75^{+1.75}_{-0.72}$ & $ \rm km\,s^{-1} $\\  
            $\phi$ & $3.09^{+1.68}_{-1.79}$ &  radians \\  
            $v_{\rm sys}$ & $88.61^{+0.42}_{-0.40}$ & $ \rm km\,s^{-1}$\\  
            \hline 
        \end{tabular}  
        
        \label{table:total}
        \vspace{13pt}
    \end{table}
\endgroup

\begingroup 
    \setlength{\tabcolsep}{9pt} 
    \renewcommand{\arraystretch}{1.5} 
    \setlength\extrarowheight{2pt}
    \begin{table}
        \centering
        \caption{ \small Summary of the model parameter statistics for the reduced population $V_1$ model. The values shown are the median values of the posterior distributions. The ranges displayed cover a $68\%$ credible interval}
        
        \begin{tabular}{ c c c }  
            Parameter & Estimate & Units\\
            \hline     
            $ \sigma_{v}$ & $0.55^{+1.33}_{-0.48}$ & $\rm km\,s^{-1}$\\
            $A$ & $0.72^{+1.38}_{-0.66}$ & $\rm km\,s^{-1} $\\  
            $\phi$ & $4.53^{+1.06}_{-3.58}$ &  radians \\  
            $v_{\rm sys}$ & $88.60^{+0.14}_{-0.08}$ & $\rm km\,s^{-1}$\\  
            \hline  
        \end{tabular}  
        
        \label{table:reduced}
        \vspace{13pt}
    \end{table}
\endgroup

\subsection{Total Population}
\label{subsec: total}

\begin{figure*}[!ht]
    \centering
    \includegraphics[width=16cm]{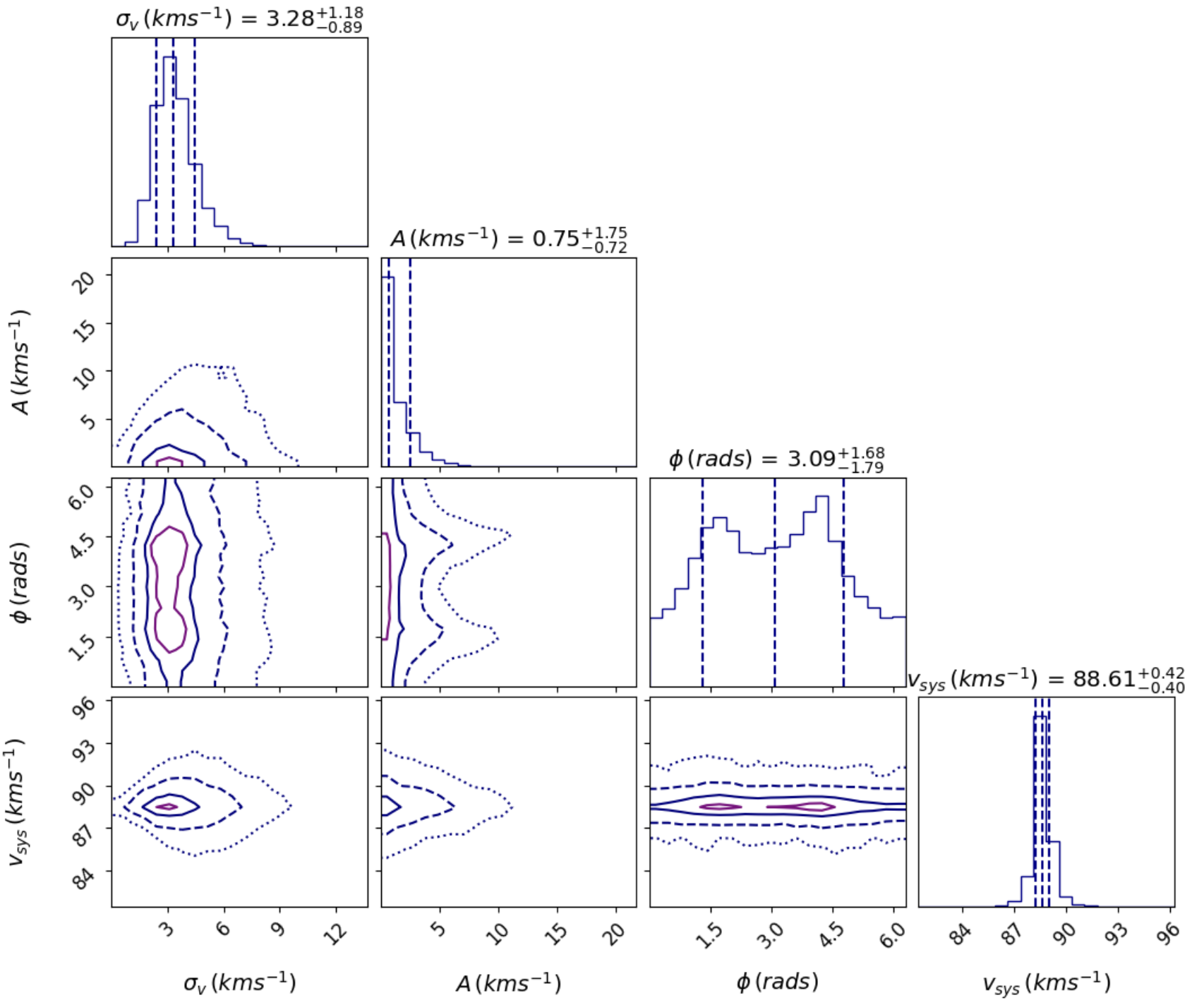}
    \caption{Posterior distributions of model parameters for the total population}
    \label{fig:total}
\end{figure*}

First, the total population of \umaiii\ is considered, with the posterior distributions of the model parameters shown in Figure \ref{fig:total}. There is a clear right skewness in distribution for $A$, with the sample $95\%$ quantile occurring at $A = 4.2 \, \rm km\,s^{-1}$. The systemic velocity and velocity dispersion are clearly bounded at $v_{\rm sys} = 88.6 \, \rm km\,s^{-1}$ and $\sigma_{v} = 3.28 \, \rm km\,s^{-1}$  respectively. There are no obvious preferences for the rotation angle, $\phi$. A summary of the model parameter statistics is shown in Table \ref{table:total}. The marginal likelihood of this model is $\ln{\mathcal{Z}} = -34.7 \pm 0.1$. The marginal likelihood for this model in the non-rotating set up (i.e. $A = 0 \, \rm km\,s^{-1}$) is $\ln{\mathcal{Z}} = -32.2 \pm 0.1$.

\subsection{Reduced Population}
\label{subsec: reduced}

\begin{figure*}[!ht]
    \centering
    \includegraphics[width=16cm]{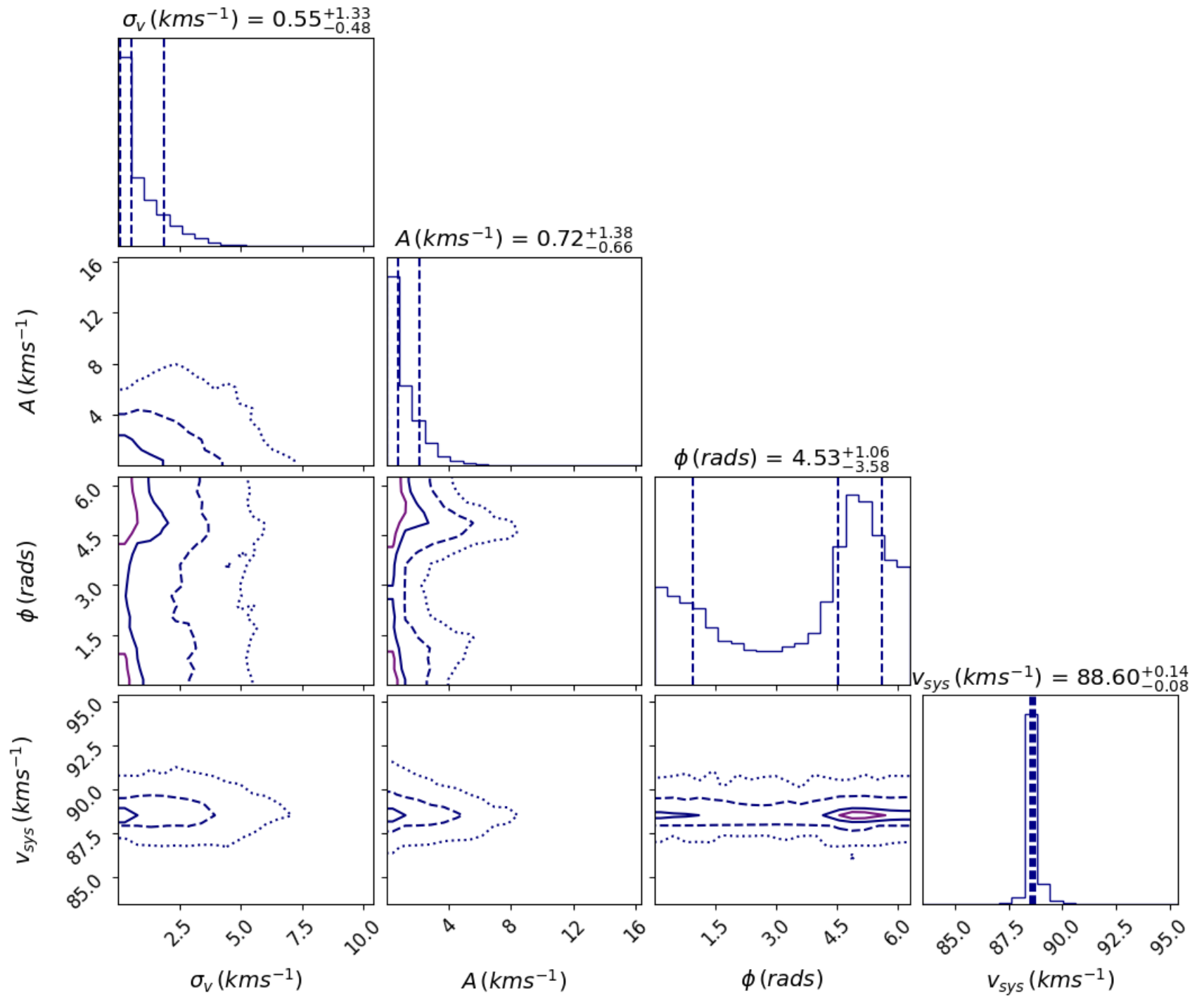}
    \caption{Posterior distributions of model parameters for the reduced population.}
    \label{fig:reduced}
\end{figure*}

\begingroup 
    \setlength{\tabcolsep}{5pt} 
    \renewcommand{\arraystretch}{1.5} 
    \setlength\extrarowheight{2pt}
    \begin{table}
        \centering
        \caption{ \small Table of marginal likelihood estimates for the Rotational and Non-rotational models for the total and reduced populations.}
        
        \begin{tabular}{ c c c }  
            Model & Full Population & Reduced Population\\
            \hline    
            Non-Rotating & $-32.2 \pm 0.1$  & $-23.8 \pm 0.1$\\
            Rotating & $-34.7 \pm 0.1$ & $-25.4 \pm 0.1$\\  
            Bayes Factor & $12.2 \pm1.3$ &  $4.9 \pm 0.5$ \\  
            \hline
        \end{tabular}  
        
        \label{table:evidence}
        \vspace{13pt}
    \end{table}
\endgroup

Member stars are now omitted from the population using the same method as \citet{2024ApJ...961...92S}, where first the largest velocity outlier is excluded, then another member is excluded due to a high S/N ratio (these are stars \#2 and \#4 in Table 3 in \citet{2024ApJ...961...92S}). The posterior distributions of the reduced $V_1$ model are shown in Figure \ref{fig:reduced}. 

Similar to the total population, the rotational amplitude exhibits a strong right skewness, with the sample $95\%$ quantile occurring at $A = 3.3 \, \rm km\,s^{-1}$. However, in contrast to the total population, the velocity dispersion has now become unresolved, with the sample $95\%$ quantile occurring at $\sigma_{v} =3.1 \, \rm km\,s^{-1}$. This sensitivity of the velocity dispersion can be attributed to the same effect seen by \citet{2024ApJ...961...92S}. The rotational amplitudes of $A = 0.75 \, \rm km\,s^{-1}$ and $A = 0.72 \, \rm km\,s^{-1}$ for the total and reduced populations respectively indicate a preference for a non-rotating model. This is shown in the $v_{\rm sys}$ vs $A$ panel in Figures \ref{fig:total} and \ref{fig:reduced}, where the peak values for both $v_{\rm sys}$ and $A$ are tightly bounded. Interestingly there is now a clearer preference in the rotation angle, but it is important to note that there are now only $9$ members in this population. A summary of the model parameter statistics is shown in Table \ref{table:reduced}. The marginal likelihood of this model is $\ln{\mathcal{Z}} = -25.4 \pm 0.1$. The marginal likelihood for this model in the non-rotating set up is $\ln{\mathcal{Z}} = -23.8 \pm 0.1$. A full comparison between models is shown in Table \ref{table:evidence}. The reason that the inferred velocity dispersions for both the total and reduced populations differ to those found by \citet{2024ApJ...961...92S} is due to the choice of the prior on $\sigma_v$, rather than the introduction of a rotational amplitude. In both scenarios the non-rotating model is favoured, with the total population displaying the stronger preference for this configuration.

\subsection{Mass Estimates}
\label{subsec: massestimates}

\begin{figure}[t!]
    \centering
    \includegraphics[width=8.5cm]{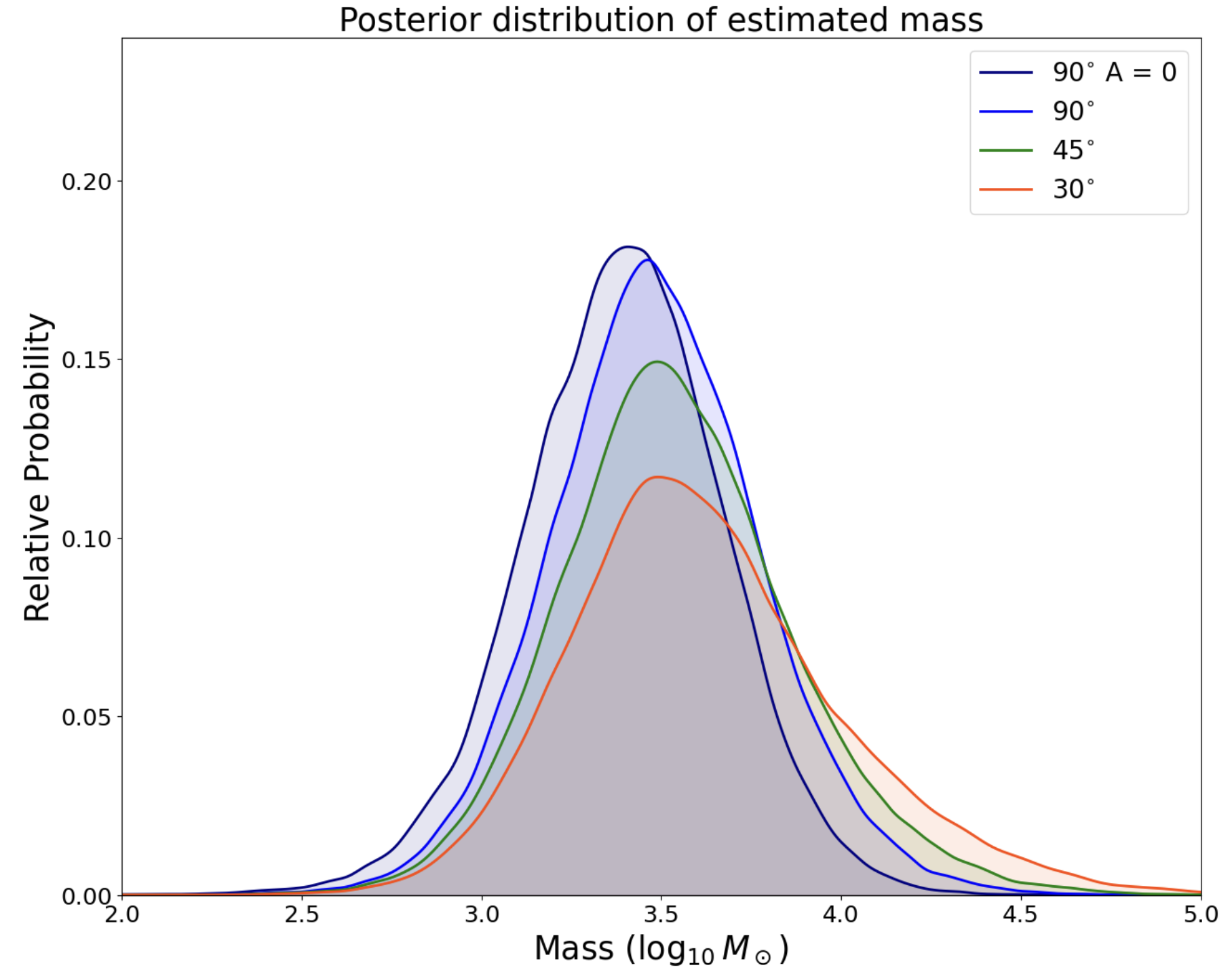}
    \caption{Posterior distribution of the estimated mass of \umaiii\ \, at inclination angles of $30^{\circ}$, $45^{\circ}$ and $90^{\circ}$. The non-rotating posterior distribution for an inclination of $90^{\circ}$ is also shown.}
    \label{fig:mass}
\end{figure}

The same mass estimator used by \citet{2020MNRAS.491L...1L} has been utilised here. All mass calculations are done using the total population. The estimated mass enclosed within a certain radius $r$ is given by:

\begin{equation}\label{eq:mass}
    M(<r) = \eta \left ( \left(\frac{v_{\rm rot}}{\sin(i)}\right)^2 + \sigma_{v}^2\right) \frac{r}{G}
\end{equation}

\noindent Where $v_{\rm rot}$ is given by Equation \ref{eq:vrot}, $\sigma_v$ is the velocity dispersion, $i$ is the inclination angle of rotation and $G$ is Newton's gravitational constant. For this investigation, the enclosed radius $r$ is set to the physical, 3D de-projected half-light radius of $r = 3 \, \rm pc$. $\eta$ is a virial parameter and is set to $1$ here. The commonly used Wolf estimator uses $\eta = 3$ \citep{2010MNRAS.406.1220W}. It is important to note that this mass calculation does not depend on $v_{\rm sys}$ and is to be seen as a lower bound estimate that doesn't include anisotropic and structural calibrations. 

To compute the mass posterior distributions, each parameter vector in the posterior sample will be used in Equation \ref{eq:mass} to obtain a corresponding estimated mass. It is important that the entire sample of $10^5$ chains is being used here, not the median values found in Tables \ref{table:total} and \ref{table:reduced}. The distribution of the estimated masses for inclination angles of $30^{\circ}$, $45^{\circ}$ and $90^{\circ} $ are shown in Figure \ref{fig:mass}.

Under the best case scenario of $30^{\circ}$ inclination (this inclination angle gives the greatest estimated rotational mass of the three chosen angles) and using a half-light luminosity of $5.7\pm1.8 \, \lsun$  \citep[from][equivalent to an absolute V-band magnitude of $+2.2$ mag, assuming a baryonic mass-to-light ratio of $ \rm M /\rm L= 1.4$]{2024ApJ...961...92S}, the lower bound mass-to-light ratio for the total population is $734.4^{+339.0}_{-176.2} \, \msun/\lsun$. Member stars are then removed following the same methodology as \citet{2024ApJ...961...92S}. For the population with a single member removed, the lower bound mass-to-light ratio becomes $123.6^{+57.0}_{-29.7} \, \msun/\lsun$. The mass-to-light ratio becomes unresolved once two members are removed from the population, which is a result consistent with \citet{2024ApJ...961...92S}. 

\section{Discussion/Conclusions}
\label{sec:conclusions}

In this paper, we examined the evidence of rotation in potentially the faintest dwarf galaxy known, \umaiii. Through a Bayesian approach, we tested a simple rotational model for both the total velocity member population as well as a reduced group. In both instances rotation was not preferred. Bayes's factors of $K = 12.2 \pm 1.3$ and $ K = 4.9 \pm 0.5$ were found upon comparison between the non-rotating and rotating models for the total and reduced populations respectively. For the total population, we see a velocity dispersion value of $\sigma_v = 3.28^{+1.18}_{-0.89} \, \rm km\,s^{-1}$, which is consistent with the value of $\sigma_v = 3.7 \, \rm km\,s^{-1}$ found by \citet{2024ApJ...961...92S}. It's important to note here that $\sigma_v = 3.28^{+1.18}_{-0.89} \, \rm km\,s^{-1}$ is within the likely range for a self-gravitating star cluster \citep{devlin2025reevaluatinguma3u1starcluster}. We observe sensitivity in the velocity dispersion to the number of members in the population, which is consistent with the work done by \citet{2024ApJ...961...92S} and \citet{2024ApJ...965...20E}. Under the better case scenario of $30^{\circ}$ inclination, we estimate the rotational mass $M_{\mathrm{\umaiii}} < 10^{5.02} \, \msun$ with $99.99\%$ probability. Then using a half-light luminosity of $5.7 \pm 1.8 \, \lsun$ (equivalent to an absolute V-band magnitude of $+2.2$ mag), we obtain a lower bound mass-to-light ratio of $734.4^{+339.0}_{-176.2} \, \msun/\lsun$ for the total population, then dropping to $123.6^{+57.0}_{-29.7} \, \msun/\lsun$ for the group without the largest velocity outlier, finally becoming unresolved once two members are removed from the population. These mass-to-light ratios are significantly smaller from those found by \citet{2024ApJ...961...92S} for the same populations ($ 6500 \, \msun/\lsun$ and $1900 \, \msun/\lsun$ respectively), however, we have not included any structural/anisotropic calibrations which leads to a factor of $3$ difference to the commonly used Wolf estimator \citep{2010MNRAS.406.1220W}.

The minimal rotation observed here indicates that \umaiii\ is unlikely to be supported by rotational pressure, with a non-rotating model $\sim5-12 \times$ more likely depending on the population chosen. Even though \umaiii\ is unlikely to be supported by rotational pressure, it may still be supported through DM pressure, as the lower bound mass-to-light ratio of $734.4^{+339.0}_{-176.2} \, \msun/\lsun$ for the total population is still well-above the needed $\sim 10 \, \msun/\lsun$ to classify a system as dark-matter dominated if binary contamination and tidal disruption are ignored. Dynamical mass-to-light ratios of $>10^3 \, \msun/\lsun$ are not unique to galaxies, however, as self-gravitating dark star clusters can reach these levels through energy ejections via black hole subsystems \citep{Rostami_Shirazi_2025}. Additionally, as highlighted here and in previous kinematic analyses done by \citet{2024ApJ...961...92S, devlin2025reevaluatinguma3u1starcluster}, all derived quantities, such as the mass-to-light ratio, are sensitive to the inclusion/exclusion of particular stars in the population. As a result, \umaiii\ still remains an ambiguous object, with both the UFD and self-gravitating star cluster scenarios still being viable. We conclude that for whichever scenario that \umaiii\ turns out to be, it is very unlikely that it is supported by rotation. In hopes to help reduce this ambiguity for future studies, continuing velocity measurements of \umaiii\ are needed.

During the final stages of the refereeing of this paper, \citet{cerny2025} published a second epoch of velocity measurements. We have undertaken preliminary analysis of these new velocities, finding that the conclusion of the lack of a significant rotational component is supported. We will present a more detailed analysis in a forthcoming contribution.

\section*{Acknowledgments}

TRA thanks the University of Sydney for a Faculty of Science Research Stipend Scholarship.

\section*{Data Availability}

Any reasonable request will be granted access to the data used in this work.

\bibliographystyle{mnras}
\bibliography{umaiii}

\end{document}